\documentclass[sigconf,nonacm=true]{acmart}
\acmConference[EASE 2025]{The 29th International Conference on Evaluation and Assessment in Software Engineering}{17–20 June, 2025}{Istanbul, Türkiye}
\usepackage[utf8]{inputenc}
\usepackage{cleveref}

\usepackage{todonotes}
\usepackage{natbib}
\usepackage{tcolorbox}
\usepackage{hyperref}
\usepackage{xcolor}
\usepackage{xspace}
\usepackage{listings}


\newcommand{\TODO}[1]{\todo[color=orange!40, inline]{\footnotesize \textbf{TODO} #1}}
\renewcommand{\TODO}[1]{{}}

\newcommand{\smalltt}[1]{{\small\texttt{#1}}}

\newcommand\ie{\emph{i.e.},\xspace}
\newcommand\eg{\emph{e.g.},\xspace}

\newcommand{\citefig}[1]{Figure~\ref{fig:#1}}
\newcommand{\citetable}[1]{Table~\ref{tab:#1}}

\newcommand{\prompt}[1]{
\begin{promptBox}
\begin{small}
  \raggedright #1
\end{small}
\end{promptBox}
}
\newcommand{\vtikz}[0]{vTikZ\xspace}

\newcommand{\AnnotatedVariants}[0]{300\xspace}

\definecolor{highlight}{rgb}{0.9,0.2,0.2}
\def\code#1{\texttt{#1}}

\definecolor{jredleft}{RGB}{255, 0, 0} 
\definecolor{jredinner}{RGB}{255, 230, 230} 
\newtcolorbox{promptBox}{
    colback=gray!10,         
    colframe=gray!30,        
    boxrule=0.3pt,           
       left=2pt, right=2pt,
    top=2pt, bottom=2pt,
    boxsep=2pt,
    width=\columnwidth
}

\title{LLM Code Customization with Visual Results: A Benchmark on TikZ}
\author{Charly Reux}
\affiliation{%
  \institution{Univ Rennes, Inria, IRISA, INSA}
  \city{Rennes}
  \country{France}}
\email{charly.reux@inria.fr}

\author{Mathieu Acher}
\affiliation{%
  \institution{Univ Rennes, Inria, CNRS, IUF, IRISA}
  \city{Rennes}
  \country{France}}
\email{mathieu.acher@irisa.fr}
\date{March 2025}

\author{Djamel Eddine Khelladi}
\affiliation{%
  \institution{Univ Rennes, Inria, CNRS, IRISA}
  \city{Rennes}
  \country{France}}
\email{djamel-eddine.khelladi@irisa.fr}

\author{Clément Quinton}
\affiliation{%
  \institution{Univ. Lille, CNRS, Inria}
  \city{Lille}
  \country{France}}
\email{clement.quinton@univ-lille.fr}

\author{Olivier Barais}
\affiliation{%
  \institution{Univ. Rennes, IRISA, Inria}
  \city{Rennes}
  \country{France}}
\email{olivier.barais@irisa.fr}

\begin{document}

\begin{abstract}
With the rise of AI-based code generation, customizing existing code out of natural language instructions to modify visual results -- such as figures or images -- has become possible, promising to reduce the need for deep programming expertise. However, even experienced developers can struggle with this task, as it requires identifying relevant code regions (feature location), generating valid code variants, and ensuring the modifications reliably align with user intent. In this paper, we introduce \vtikz, the first benchmark designed to evaluate the ability of Large Language Models (LLMs) to customize code while preserving coherent visual outcomes. Our benchmark consists of carefully curated \vtikz editing scenarios, parameterized ground truths, and a reviewing tool that leverages visual feedback to assess correctness. Empirical evaluation with state-of-the-art LLMs shows that existing solutions struggle to reliably modify code in alignment with visual intent, highlighting a gap in current AI-assisted code editing approaches. We argue that \vtikz opens new research directions for integrating LLMs with visual feedback mechanisms to improve code customization tasks in various domains beyond TikZ, including image processing, art creation, Web design, and 3D modeling. 
\end{abstract}

\maketitle

\vspace*{-2mm}
\section{Introduction}

Customizing code that produces visual results is a general and fundamental problem, with numerous applications in image processing (\eg SVG manipulation), digital art creation (\eg p5.js sketches), Web development (\eg HTML/CSS layouts with backend interactions), and 3D modeling (\eg Blender scripting, SCAD programs).
In such contexts, modifications have to be made at the code level in order to achieve a specific visual change, such as adjusting an object's shape, changing colors, or adding elements. 
Rapid progress has been achieved in the field of Large Language Model (LLM)-assisted software engineering~\cite{hou_large_2024}. 
With the emergence of AI-based code generation, customizing code through natural language instructions (prompts) is more and more investigated, offering new possibilities for end-user programming and even for developers who need to onboard or interact with codebases~\cite{acher_programming_2023,liu_empirical_2024,liu_large_2024}. 
However, modifying code with visual results remains a complex task, even for experienced developers, as it requires identifying the relevant code regions (feature location), generating valid code variants, and ensuring that modifications align with user intent while maintaining reliable and consistent visual results.

\begin{figure*}
    \centering
    \includegraphics[width=\linewidth]{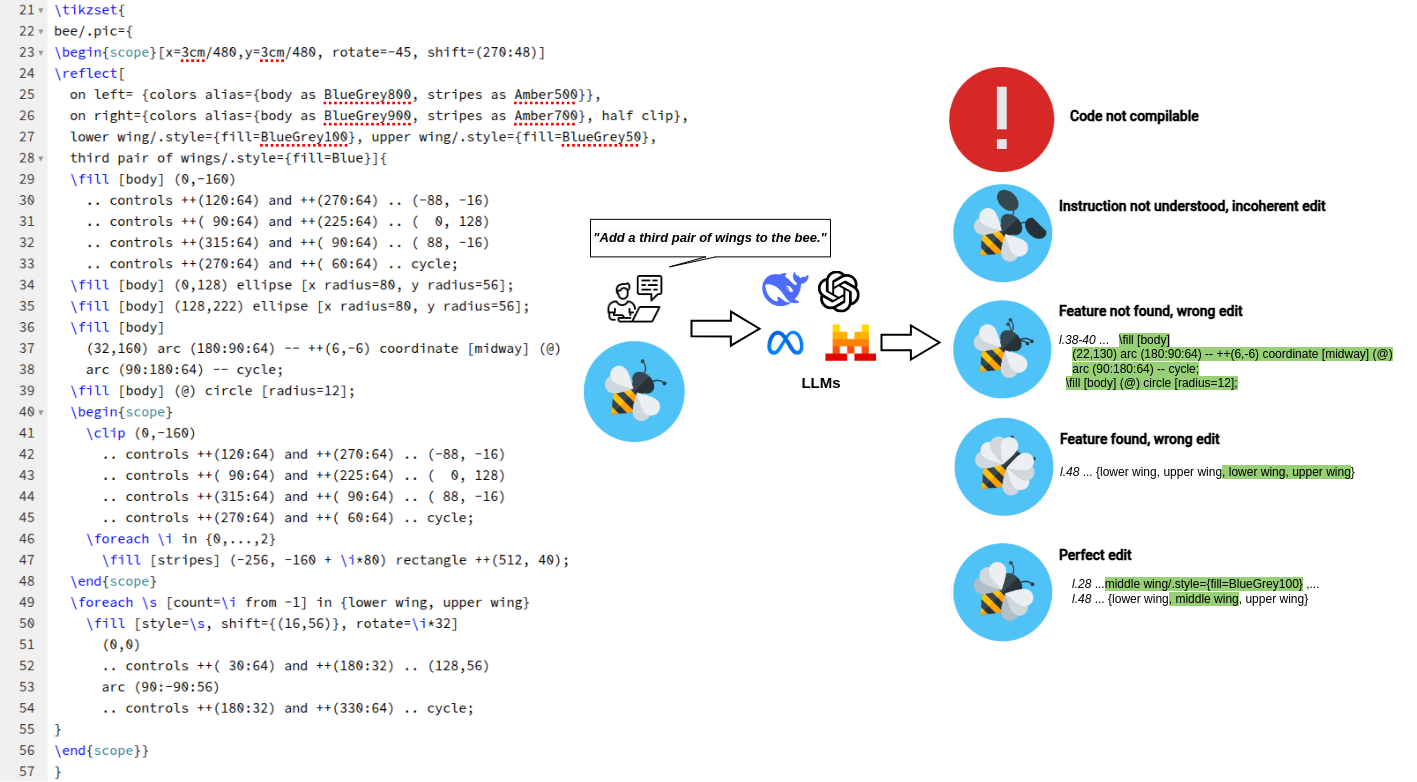}
   \caption{
       Contextualized example of the benchmark task. 
   }
    \label{fig:context}
\end{figure*}

Users can typically prompt an LLM with high-level instructions, detailing the intended visuals supposed to be produced by a program, rather than the technical aspects of the code itself. It is useful for instance when tweaking interfaces, manipulating diagrams, or customizing images or 3D models coming from tools, scripts, or programs. 
 To illustrate, consider a TikZ script that generates an image of a bee. A user may request an LLM to "add a third pair of wings", a seemingly simple instruction that requires multiple non-trivial steps: understanding the intent, identifying the relevant code, and applying the correct modification. Unlike humans, who can mentally simulate the visual effect of code changes, LLMs often struggle with these steps, behaving like a blind system guessing modifications without direct feedback from the rendered output.
To systematically study this problem, we identify three key challenges:
\begin{itemize}
    \item Feature location: Identifying the correct code segments that contribute to the visual elements requiring modification. Many visual programming languages, such as TikZ, lack explicit mappings between rendered elements and the corresponding source code.
    \item Code customization and variant synthesis: Generating alternative versions of the code that satisfy the modification request while ensuring syntactic correctness and maintaining consistency across multiple lines of code.
    \item Visual result validation: Ensuring that the generated output aligns with the intended change, remains visually coherent, and does not introduce unintended modifications to other elements in the figure.
\end{itemize}
Evaluating LLMs on tasks involving code customization and multimodality is promising, as it addresses both cross-modal consistency and program behavior prediction. As shown in \Cref{fig:context}, LLMs can sometimes customize code correctly. However, systematically quantifying whether customized code meets an instruction remains a challenge, as no true ground truth exists for comparison. Our observation is that multiple edits can produce correct outputs, but vision-based oracles can only approximate human preferences without certainty.

To explore these challenges, we introduce \vtikz, the first benchmark designed to evaluate the ability of LLMs to customize code while preserving coherent visual results. \vtikz consists of 100 manually curated TikZ editing scenarios, each requiring code modifications to achieve a specified visual change. Unlike existing benchmarks, which focus either on textual code edits (\eg bug fixes, refactorings) or visual-based interactions (\eg autonomous Web agents), \vtikz explicitly evaluates code customization in the presence of a visual output, bridging the gap between textual and visual modalities.
We have carefully designed parameterized ground truths to account for multiple valid solutions, avoiding the unfair penalization of LLM-generated variants that still produce correct outputs. Additionally, we developed a reviewing tool to assess correctness based on the generated images. Using this tool, we have manually annotated \AnnotatedVariants LLM-generated variants, that enabled a fine-grained validation of our ground-truths. 
Furthermore, we perform an evaluation of different LLMs. Results show that LLMs struggle to customize code, validating the design of \vtikz and calling to LLM-based solutions augmented with richer tools, modalities, or feedback.

To the best of our knowledge, no existing benchmark explicitly tackles LLM-driven code customization with visual validation. On the one hand, several software engineering (SE) benchmarks evaluate LLMs on tasks, such as bug fixing, refactoring, and feature addition, but these tasks focus solely on textual modifications without considering visual outcomes~\cite{jimenez_swe-bench_2024,aleithan_swe-bench_2024,liang_can_2024,labash_res-q_2024}. On the other hand, some benchmarks assess visual-based AI agents, such as autonomous Web agents, which manipulate visual content, but they do not involve code customization. 
 In between, there have been a few studies on LLM-based TikZ generation, where models are prompted to create figures from scratch based on textual descriptions~\cite{belouadi_detikzify_2024,wei_words_2024}. However, these approaches differ significantly from our customization scenario, where the challenge lies in modifying existing code while preserving intent. In the (grey) literature, TikZ is often used informally to evaluate LLMs with challenges like "draw a unicorn"~\cite{bubeck_sparks_2023,willison_notes_2025}. However, these evaluations lack systematic benchmarking. \vtikz extends this direction by providing a rigorous benchmark with structured evaluation metrics and a focus on modifying existing code rather than generating from scratch.

In this work, we introduce \vtikz, a benchmark designed to evaluate LLMs in customizing code with visual results. Our contributions are as follows:
\vspace*{-2mm}
\begin{itemize}
    \item A dataset of 100 manually curated TikZ customization tasks, covering a variety of modifications (e.g., adding, removing, resizing, and repositioning elements) to assess LLMs' ability to modify existing graphical code.
    \item A parameterized ground truth framework, acknowledging that multiple code variants may correctly implement a given visual modification, preventing unfair penalization of LLM-generated outputs.
    \item A visual reviewing tool that enables automated and human-in-the-loop evaluation of generated visual results. Using this tool, we collected \AnnotatedVariants data points, refining our evaluation framework and validating our ground-truth design.
    \item A systematic evaluation of LLMs on vTikZ, analyzing their strengths and limitations in feature location, code modification, and visual consistency. Our results show that existing models struggle with code customization, reinforcing the need for hybrid approaches integrating additional tools, multimodal feedback, or iterative validation mechanisms.
    \item An extensible benchmark that provides a structured methodology for evaluating LLM-driven code customization with visual outputs, paving the way for future research in multimodal AI-assisted programming.
\end{itemize}

The \vtikz dataset, the data resulting from the evaluation, and the manually annotated data are available online\footnote{\url{https://huggingface.co/datasets/CharlyR/vtikz}}, along with the benchmark's code\footnote{\url{https://github.com/IV2C/VTikZ}}.


\section{Background and Motivation}

In this section, we define the terminology used in the remainder of the paper and we motivate the need for a benchmark on TikZ with an illustrative example.

\subsection{Background and Terminology}

The work and the benchmark presented in this paper revolve around three main notions, namely TikZ, variants, and patches.
\textit{TikZ} is a tool to create graphic elements in \LaTeX. It is used mainly for scientific diagrams, although users have found other use cases, notably for drawings and cartoon characters. \label{def:tikz}
We define a \textit{variant} as a piece of TikZ code on which a \textit{patch} 
has been applied.
A patch is defined as the edits (\ie changes) between two versions of a code represented in the unidiff format, it is the same format used by Git when displaying changes between commits. 





\subsection{Motivating Example}

To illustrate our work, we consider an example consisting in a TikZ code snippet that generates an image of a bee with two wings on each side, as shown on the left side of \citefig{context}. 
Although the bee is a simple drawing, the underlying TikZ code is not trivial to understand and manipulate, especially for developers without prior expertise. 
Let us suppose a developer wants to customize the bee's design by modifying the TikZ code. 
One way to proceed is by leveraging an LLM, providing it the existing TikZ code together with a change instruction, such as ``\textit{Add a third pair of wings to the bee}''. 
The instruction, though simple in its intent, requires that the LLM perform multiple non-trivial sub-tasks. First, it must \textit{understand the instruction}: What does "a third pair of wings to the bee" mean? While a human perfectly gets the intent, this is not necessarily the case for an LLM.
Next, it must \textit{identify the relevant feature}: Where are the wings defined within the TikZ code? Assuming the LLM properly interprets the instruction, it must locate the corresponding code fragment that requires modification.
Finally, it must \textit{perform the correct modification}: What changes need to be made in the code? Once the relevant lines of code are identified, the LLM must determine the appropriate additions, deletions, or modifications to reach the intended result.

Possible mistakes an LLM can make in these sub-tasks are illustrated on the right side of \Cref{fig:context}. First, the worst-case scenario is generating code that fails to compile. Second, the LLM might misinterpret the instruction, leading it to modify an unintended feature, such as increasing the size of the antennas instead of adding wings. Third, it may fail to identify the relevant feature, which could result in unintended modifications like adding an extra pair of antennas. Fourth, even when the relevant feature is correctly identified, the LLM could make incorrect modifications, for example by introducing two additional pairs of wings instead of just one. Fifth and finally, the LLM might overlook contextual subtleties; even when the edit is correct, details such as ensuring the shading of the newly added wings remains consistent with the original design must still be addressed.

These cases show how challenging customizing TikZ code is.
However, even after completing these tasks, another challenge remains: the modifications made by an LLM may introduce subtle deviations from a theoretically perfect solution. 
For example, the newly added bee wings might exhibit slight color variations or be misaligned. 
Such cases complicate the assessment of modifications and make the definition of a precise oracle for evaluating correctness challenging.
Addressing this issue is an objective of our work. 


\section{Dataset}

This section details the dataset used in our benchmark, which we refer to as the \emph{\vtikz} dataset. 
It consists of a total of 100 human-made TikZ variants, each derived from an original TikZ code collected from the internet, along with the corresponding instruction describing the edit used to create the variant. 
The following sections describe the dataset's selection, curation process, format, and content.

\subsection{Selection and Curation of TikZ code}

\begin{figure}
    \centering
    \includegraphics[width=\linewidth]{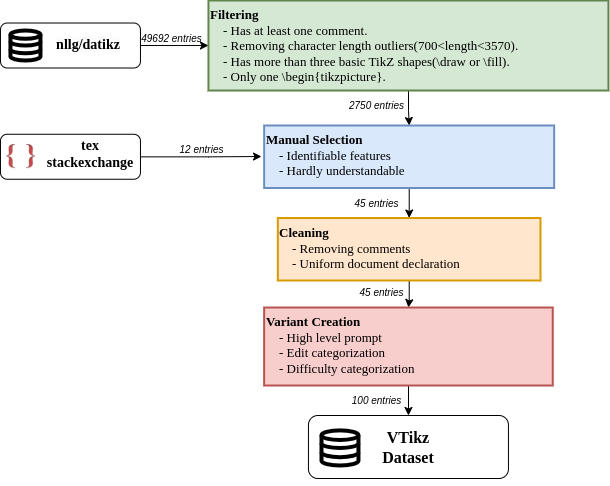}
    \caption{
        \vtikz dataset curation overview
    }
    \label{fig:curation}
\end{figure}

This section details the protocol we followed to select and curate the set of TikZ codes. 
The creation of such a dataset involved identifying TikZ code examples that exhibited both \textit{identifiable features} and \textit{inherent difficulty}, making them challenging for end-users to understand at first sight. Such examples illustrate the complexity of TikZ while maintaining a well-defined structure. 
They also support high-level prompts specifying feature edits in the diagram, \eg \textit{"Make the eye of the dog red."}.
\Cref{fig:curation} shows the curation protocol we followed.
We collected data from two repositories, \ie 
animal drawings from a post on the \LaTeX{}  StackExchange\footnote{\url{https://tex.stackexchange.com/a/414059}}, and scientific diagrams from the nllg/datikz\footnote{\url{https://huggingface.co/datasets/nllg/datikz}} dataset.
To ensure the selected TikZ code was well-suited for our benchmark's customization task, we applied the following filtering criteria to the nllg/datikz dataset: 

\begin{itemize}
    \item \textbf{At least one comment}, as comments were more likely to indicate higher-level attributes.
    \item \textbf{No character length outliers}, by removing codes with fewer than 700 characters or more than 3570. Computed with\\
    \(\left[ Q_1 - 1.5 \times \text{IQR},\ Q_3 + 1.5 \times \text{IQR} \right]\).
    
    \item \textbf{More than 3 basic TikZ shapes}, by limiting to codes primarily composed of TikZ elements (with three commands \code{\textbackslash fill} or \code{\textbackslash draw}).
    
    \item \textbf{Only one \code{\textbackslash begin\{tikzpicture\}} command}, ensuring a single TikZ diagram is generated.
\end{itemize}

We developed these criteria through an initial manual exploration of the data, iteratively refining them based on the structural characteristics of the analyzed TikZ code. This process resulted in 12 codes from StackExchange and 2750 from the nllg/datikz dataset. The remaining dataset creation was conducted by one annotator, who was responsible for creating the customizations. To mitigate the bias induced by relying on a single annotator, the created data was reviewed and discussed with four software engineering experts.
While this could introduce a new source of bias, the simplicity of the customizations, combined with the ability to visually compare the results, ensures that there is little to no ambiguity in the potential solutions, making the solutions straightforward to review.
\TODO{paragraph has been reformulated for "No clear selection criteria are given for choosing code samples"} Through a random exploration of the remaining 2750 codes, we manually selected 35 codes that were more prone to being customized, and handpicked 10 codes from the 12 initial ones from stackexchange, for a total of 45 initial codes.
For each TikZ code, we manually created one or more variants, with different types of changes and difficulty, leading to a total of 100 variants.

\subsection{Format}
\label{sec:dataset_format}

The format of our dataset includes both human-made and automatically computed features, as detailed in \Cref{tab:features}. 
The human-made features for each customization consist of:  the \textit{original code} in TikZ, a list of \textit{parameterized solution codes},  the \textit{instruction},  the \textit{perceived difficulty}, the \textit{description of the resulting modification},  the \textit{type of diagram} and the \textit{type of modification}.  

\begin{table}
    \centering
    \caption{
        Human-made and computed features in the \vtikz Dataset.
    }
    \fontsize{7}{8}\selectfont
\begin{tabular}{@{}p{3cm}p{5.3cm}@{}}
\toprule
\multicolumn{2}{l}{\textbf{Human-made Features}} \\
\midrule
Original Code       & Source code provided by humans. \\
Parameterized Solution Code             & A list of code parameterized to encompass valid solutions. \\
Instruction         & Edit instruction. \\
Perceived Difficulty& Subjective difficulty of the edit for each possible variant code. \\
Result Description  & Description of the edit outcome. \\
Modification Type      & Type of modification applied by the prompt(update, add, remove). \\
Type      & Type of the original diagram, either "scientific" or "animal". \\
\midrule
\multicolumn{2}{l}{\textbf{Computed Features}} \\
\midrule
Perfect Variants             & List of theoretically perfect variants, computed from the parameterized solution code. \\
Patch               & Unidiff patches (without context) between original and theoretically perfect variants. \\
Image Input         & Image rendered from the original code. \\
Images Solution      & List of images rendered from the theoretically perfect variants. \\
AST Difficulty      & Tree Edit Distance between original and variant \cite{zhang_simple_1989}. \\
\bottomrule
\end{tabular}

    \label{tab:features}
\end{table}

From these human-made features, we derive the following computed features:  
\begin{itemize}
    \item \textbf{Perfect Variants}: A list of theoretically perfect \textit{variants} that fully implement the instruction. 
    
    \item \textbf{Patch}: A list of string representations of the edits in Unidiff format (without context), computed by comparing the original TikZ code with its modified variants.
    
    \item \textbf{Image Input}: A rendered image of the original TikZ code.  
    
    \item \textbf{Images Solution}: A list of rendered images generated from the theoretically perfect \textit{variants}' code.  
    
    \item \textbf{AST Difficulty}: The Tree Edit Distance between the original and variants TikZ codes, computed using the algorithm proposed by \citet{zhang_simple_1989}.
    
\end{itemize}

These features are used in the evaluation and analysis phase of our benchmark, which we both further detail in \Cref{sec:benchmark} and \ref{sec:analysis}.

\label{sec:param_code}

As mentioned earlier, multiple solutions can be valid. For example, when instructing an LLM to change a drawing element to red, various shades of red are acceptable. Similarly, when asking to move a drawing to the right, the distance should be parameterizable, as multiple shifts can be valid. To cover as many valid solutions as possible in our evaluation, we introduce four parameterization statements that can be added to TikZ code:

\begin{itemize}
    \addtolength{\leftskip}{-15pt}
    \item \smalltt{\S range(lower, higher, default)}: Specifies a valid numerical range between \emph{lower} and \emph{higher}, with a \emph{default} value. Multiple valid radius sizes can thus be defined as follows:\\ 
\smalltt{\textbackslash fill [Red600] (56, 0) circle [radius=\textcolor{highlight}{\S range(10, 30, 20)}];}

\item \smalltt{\S rangei(value, interval)}: Functions similarly to \smalltt{\S range} but defines a range using an interval around a central value, equivalent to \smalltt{\S range(interval-value, interval+value, value)}.\\
The previous radius can be parameterized as follows:\\
\smalltt{\textbackslash fill [Red600] (56, 0) circle [radius=\textcolor{highlight}{\S rangei(20,10)}];}

\item \smalltt{\S choice([A,B,C,...], default)}: Specifies a list of valid values $A, B, C, ...$.\\ 
\smalltt{\textbackslash fill [Red\textcolor{highlight}{\S choice([600,700],600)}] (56, 0) circle [radius=20];}

\item \smalltt{\S def(value)}: Defines a variable name. 
Since different TikZ codes may use distinct variable names while remaining functionally equivalent, any code with a different variable name at \smalltt{\S def} is considered identical as long as the variable is used in the same locations. A variable name can be defined as follows:\\
\smalltt{\textbackslash definecolor\{\textcolor{highlight}{\S def(myred)}\}\{rgb\}\{0.9,0.2,0.2\}}\\
\smalltt{\textbackslash fill [myred] (56, 0) circle [radius=20];}
\end{itemize}

During the creation of the dataset, the default values defined above are used to generate the theoretically perfect variant.

However, parameterized code alone cannot encompass every solution.
To address this, we provide multiple parameterized variants that implement the same solution in different alternatives. For instance, instead of simply changing a color to red, TikZ allows defining a custom color using \smalltt{\textbackslash definecolor} and reusing it later. 
Since a single parameterization cannot cover both approaches, we can include multiple variants: one applying the red color and another defining and referencing a custom color.
\subsection{Descriptive Analysis}

\begin{table}
    \centering
    \caption{
         Dataset Summary
    }
    \resizebox{\linewidth}{!}{%
\begin{tabular}{ll|ll}
\toprule

\textbf{Edit Categories}& & \textbf{Difficulty Levels}&\\
Scientific Edits      & 50   &  Easy Edits    &   41        \\
Animals Edits         & 50   &  Medium Edits                &   36      \\
                       &      &    Hard Edits              &   23      \\
                       && Avg. AST Difficulty          & 20.42 (1–160) \\
\addlinespace
\textbf{Dataset Metrics}&&\textbf{Code Metrics}&\\
Number of Codes       & 45   &    Avg. Lines of Code           & 63 (25–104)            \\
Avg. Edits per Code   & 2.22 &    Avg. Characters              & 2167 (918–3323) \\
\addlinespace
\textbf{Edit Types}&&&\\
Add                   & 25   &         &           \\
Remove                & 13   & & \\
Update                & 62   & &\\
\bottomrule
\end{tabular}%
}

    \label{tab:dataset_analysis}
\end{table}

We now present an analysis of the dataset (\Cref{tab:dataset_analysis}), demonstrating its scope and diversity through statistics, including the number of edits, lines of code, difficulty levels, and types of modifications.

Each customization in the dataset falls into one of two main categories-\emph{Scientific} or \emph{Animals}-depending on the source of the input code, with 50 customizations in each category. 
The 45 initial codes in the \vtikz dataset contain, on average, 2.22 customizations each.
These customizations are classified into three types, \textit{add, remove, or update}, with respective counts of 25, 13, and 62. These customizations are mainly structural, involving layout, colors, sizes, and shapes.  

In total, the dataset contains in 41 easy, 36 medium, and 23 hard customizations, classified by the "perceived difficulty" feature.
The computed AST difficulty averages 20.42, ranging from a minimum of 1 to a maximum of 160. 
The input codes have an average of 63 lines of code and 2,167 characters, with lengths varying between 25 and 104 lines of code and character counts ranging from 918 to 3,323. 
Although most customizations may be considered straightforward for an experienced user, editing TikZ code presents two major challenges. First, the model must accurately \emph{identify relevant feature(s)} based on the user's instructions, a non-trivial task given the complexity and abstract nature of TikZ. 
Second, after correctly locating relevant lines of code, the LLM must \emph{perform the correct modification}, ensuring that changes do not disrupt the interdependent structure of the code.
As we will demonstrate, LLMs struggle even with seemingly simple customizations. 

\section{Benchmark}
\label{sec:benchmark}

Evaluating the code-editing capabilities of LLMs plays a major role in the Software Engineering community, ensuring among other things, the reliability, quality, and compliance of generated code.
In this section, we present the \textit{\vtikz} benchmark, in which LLMs are prompted to generate TikZ code variants based on a given instruction. Through this task, we assess the reliability of LLMs and evaluate to which extent the generated code variants comply with the given instruction.

In the remainder of the section, we first present the evaluation metrics, followed by an explanation of the task, and finally, we provide technical details regarding the benchmark.

\subsection{Evaluation metrics}
\label{sec:emetrics}
We employ five metrics to comprehensively assess the quality of a customized code variant compared to the reference solution.

\paragraph{Primary Metrics}
\begin{itemize}
  \item \textbf{CompileMetric}:
  A binary metric (0 or 100), indicating whether the generated code compiles successfully.
  \item \textbf{LocationMetric}: 
  A binary metric (0 or 100), indicating whether 100\% of the lines of a patch was edited.
  \item \textbf{SuccessCustomizationMetric}: 
  A binary metric (0 or 100), evaluating whether the generated code matches a parameterized solution or produces an exact image match.
\end{itemize}

\paragraph{Complementary Metrics}
\begin{itemize}
  \item \textbf{SimilarityMetric}: Measures similarity between patches using CrystalBleu~\cite{eghbali_crystalbleu_2023}.
  \item \textbf{LineMetric}: Computes the percentage of correctly edited lines, defined as: 
  \[
  \text{LineMetric} = \frac{100\times\text{Number of correctly edited lines}}{\text{Total number of edited lines in the reference solution}}
  \]
\end{itemize}

\subsection{Task}

Given an input TikZ code \( c \), an instruction \( p \), and a rendered image \( R(c) \) from the original code \( c \) (optionally provided as input to the model), the \( LLM \) generates a variant \( v \)
\[
v = LLM(c, p, R(c)),
\]

Because of the probabilistic nature of Large Language Models, we evaluate their performance using Best-Of-N sampling, generating \( N \) variants for each code-instruction pair \( c,p \) while varying the temperature \( t \):
\[
v_i = LLM^t(c, p, R(c)) \quad \text{for } i = 1, \dots, N.
\]

Given a single generated variant \( v_a \), we compare it against a set of \( m \) reference solutions, each represented as a tuple of a template \( t_i \) and a theoretically perfect variant \( r_i \): \(\{ (t_i, r_i) \}_{i=0}^m\).
To determine the most relevant reference tuple for comparison, we compute each metric \( M_d \) as follows:
\[
M^i_d = \text{metric}(v_a, (t_i,r_i)) \quad \text{for} \quad i = 0, 1, 2, \dots, m
\]
We then rank the reference solutions in descending order based on their metric scores (see \Cref{sec:emetrics}), selecting the tuple that yields the highest scores. 
The ranking relies on the following order of priority applied to the primary metrics: SuccessCustomizationMetric, LocationMetric, LineMetric, SimilarityMetric, CompileMetric.
For this variant \( v_a \), we then only consider the metrics' scores for the tuple that yields the best scores.
Finally, for each variant \( v_i \) in the \(N\) tries, we apply the same ranking process to select the best-performing variant. 

Following this process, we selected among the \(N\) generated variants the one that yields the best scores by comparing each with each possible solution.

\subsection{Technical Details}
The benchmark is designed to evaluate both LLMs and Large Multimodal Models (LMMs) in a comparable manner. 
In addition to textual inputs, LMMs can also process images, making them suitable for TikZ-based image modifications.
Both model types leverage system prompts, ensuring a better alignment with the benchmark objectives. 
LLM and LMM system prompts are structured as follows:

\prompt{

\textit{You are an expert coding assistant specialized in modifying file contents based on instructions.\\
Given an instruction and file content, respond only with the updated file's full content, ensuring it is entirely enclosed between code tags like this\\
```\\
content\\
'''\\
Provide no additional text or explanations beyond the code tags.}
}

For LMMs, when the image is also provided, the system prompt is modified to include information about the provided image:
\prompt{\textit{[...] Given an instruction, file content, and the image that the current file creates, [...]
}}

All models were evaluated through API inference providers, using the temperature setting provided in \citetable{avg_metric} and \citefig{human_annotation}.

\subsection{Benchmark Refinement and Validation}
\label{subsec:refinement}

\begin{figure}
    \centering
    \fbox{\includegraphics[width=0.9\linewidth]{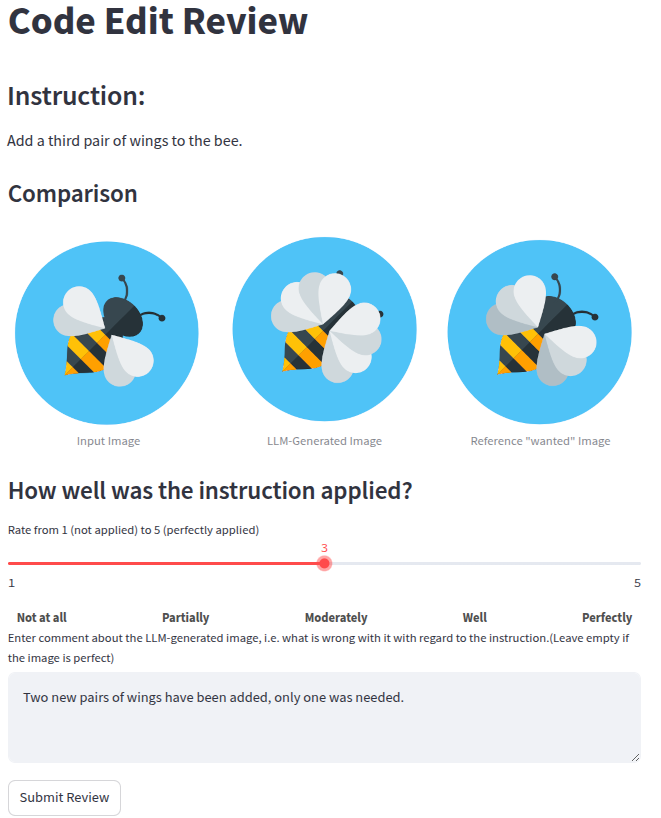}}
            \vspace*{-0.4cm}
    \caption{
        Annotation interface
    }
    \label{fig:human_annotation}
\end{figure}

We updated our benchmark in two phases: first by refining the evaluation metrics, then by validating and expanding the set of ground truths through human feedback.

The initial version focused on a limited set of text-to-text and image-to-image metrics applied to a reduced Large Language Model set. LLM-generated variants were scored on a five-point scale based on visual inspection via an early version of our annotation interface (see \Cref{fig:human_annotation}). This led to the classification of 745 variants and revealed weaknesses in metric coverage—particularly in identifying valid solutions overlooked by automated measures. These findings motivated the introduction of a \emph{parameterized solution framework}, allowing multiple valid solutions per task and better reflecting the structured nature of code edits.

\TODO{replaced "Annotators" with "One annotator" for "Who did the annotations? one or multiple annotators? if one how you mitigated bias? if multiple how did you manage disagreements?  ethics information is missing if external annotators were used."}
Building on this framework, we conducted a second evaluation using a redesigned annotation interface (also shown in \Cref{fig:human_annotation}). One annotator provided both numerical scores and qualitative feedback for each variant. Out of all generated outputs—selected and unselected across multiple attempts—2,525 were compilable. From these, we created a new dataset with \AnnotatedVariants{} annotated instances.

\TODO{One reviewer said "The manual validation results are unclear. The paper says: "This annotation process proved instrumental in validating and refining the benchmark..." but it does not say how many examples were verified or ...", I think he did not understand that 13 out of the 300 instance were right solutions made by the LLM. I merged the last two sentences.}
Each entry includes two new fields: \texttt{human\_score}, reflecting correctness and completeness, and \texttt{human\_comment}, capturing specific strengths or errors. This annotation process proved instrumental in validating and refining the benchmark, as it surfaced 13 additional correct solutions that were previously missed. 


\section{Results and Analysis}
\label{sec:analysis}

In this section, we evaluate state-of-the-art LLMs on our constructed dataset using the defined metrics, producing a benchmark that we then analyze.
Our benchmark includes five recently released open-source LLMs, namely DeepSeek-R1-Distill-Llama-70B,
Llama-3.1-8B, Llama-3.3-70B, Llama-3-70B, all in their instruct versions, and one closed source model, \ie GPT-4o(2024-08-06), evaluated both with and without an image input.
To ensure a fair comparison, we selected a temperature of 0.7 for most evaluations, as this is a commonly used default parameter among inference providers, mimicking the typical user experience.
Our analysis aims to answer the following  research question:
\emph{To what extent can LLMs generate successful code customizations?}

    

\subsection{Variant Classification}
\begin{figure*}
    \centering
    \includegraphics[width=9\linewidth/10]{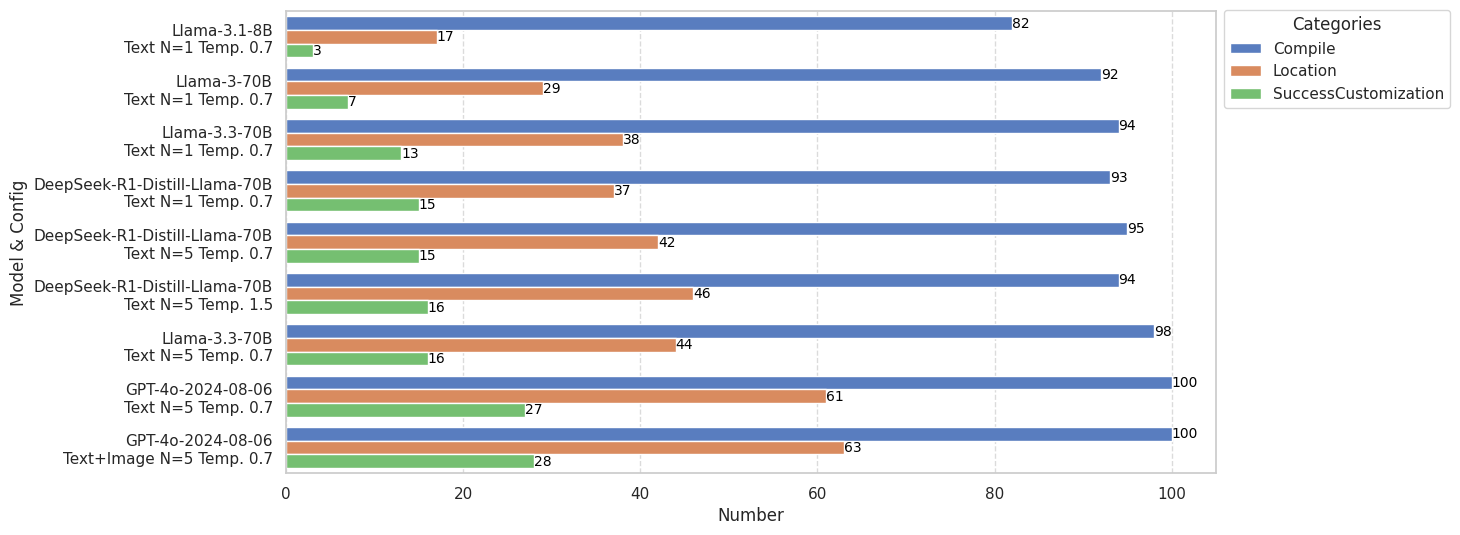}
    \caption{
        Classification of Variants using the Compile, Location, and SuccessCustomization metrics
    }
    \label{fig:variant_classif}
\end{figure*}

To assess how well LLMs can generate successful code customizations, we classified the variants using the Compile, Location, and SuccessCustomization metrics, as shown in \Cref{fig:variant_classif}.

The Compile Metric gives promising results, especially for the GPT-4o at \(k=5\), the only one with an average score of 100. Other models are close behind, the furthest being llama-3.1-8b, achieving 82. 
For the Location Metric, GPT-4o manages to edit the right lines on average 60\% of the time, while the other models lack behind with 45\% for DeepSeek-R1-Distill-Llama-70B at \(k=5\) with a temperature of 1.5. 
The SuccessCustomizationMetric only goes up to an average of 28 with GPT-4o, the Llama-3.3-70B and DeepSeek-R1-Distill-Llama-70B models achieve closer results going from 13 to 16.

\subsection{Variants Similarity To Solution}

\begin{table}
    \centering
    \caption{
         Average similarity and line scores by configuration
    }
    \resizebox{\linewidth}{!}{
\begin{tabular}{llcccc}
\toprule
Model & Modality & N & Temp. & Similarity & Line \\
\midrule
Llama-3.1-8B & Text & 1 & 0.7 & 18.2 & 31.4 \\
Mixtral-8x7B & Text & 1 & 0.7 & 17.4 & 30.7 \\
Llama-3-70B & Text & 1 & 0.7 & 29.3 & 45.6 \\
Llama-3.3-70B & Text & 1 & 0.7 & 37.9 & 54.5 \\
DeepSeek-R1-Distill-Llama-70B & Text & 1 & 0.7 & 35.8 & 48.1 \\
DeepSeek-R1-Distill-Llama-70B & Text & 5 & 0.7 & 39.0 & 58.4 \\
DeepSeek-R1-Distill-Llama-70B & Text & 5 & 1.5 & 36.9 & 57.8 \\
Llama-3.3-70B & Text & 5 & 0.7 & 41.7 & 60.6 \\
GPT-4o-2024-08-06 & Text & 5 & 0.7 & 55.7 & 75.0 \\
GPT-4o-2024-08-06 & Text+Image & 5 & 0.7 & 57.0 & 75.3 \\
\bottomrule
\end{tabular}
}

    \label{tab:avg_metric}
\end{table}

The classification from the previous section defines which variant constitutes a solution. In this section, we evaluate how close a variant is to a solution by presenting and analyzing the similarity and line scores.

The scores are presented for each model in \Cref{tab:avg_metric}, along with the number of \(k\) attempts and temperature settings. The Modalities used are either Text only(T) or Text and Image(T+I), which only applies to GPT-4o.
We observe that the Similarity Metric --computed using the CrystalBLEU score between the two patches-- exhibits a similarity of 57\% for GPT-4o, significantly outperforming all other models. On the lower end, the Mixtral-8x7B and Llama-3.1-8B models achieve similarity scores around 18\%.

Regarding the Line Metric, which measures the percentage of correctly edited lines, GPT-4o achieves an average accuracy of 75.3\%, while other models under-perform, with the closest being Llama-3.3-70B at 60.6\%.

\subsection{Observations and Analysis}

Although the exact size of the GPT-4o model is not publicly available, the scores measured with our benchmark show an upward trend with the number of parameters of the models. 
Larger models tend to perform better, particularly newer models like Llama-3.3-70B which achieves double the SuccessCustomization metric score of Llama-3-70B despite having the same number of parameters.

Surprisingly, the reasoning-focused model DeepSeek-R1-Distill-Llama-70B exhibits similar results to its non-reasoning counterpart,
suggesting that reasoning capabilities do not significantly enhance TikZ editing performance.

Giving the input image to the GPT-4o model resulted in better scores overall, and the number of successful edits decreased marginally (by only one) when solely relying on code inputs. This suggests a potential lack of multimodal consistency, at least in the context of TikZ diagrams.

We observe that the SuccessCustomization metric unfortunately varies from 3\% to 28\% of the solutions, which emphasizes the current limits of LLMs in customizing TikZ code. Regarding the Location and Compile metrics, they vary respectively from 17\% to 63\%  and from 82\% to 100\%. This shows nonetheless the existing potential in succeeding in such as task, but will require a more innovative approach to ensure a more reliable and successful customization. 
Overall, even with the best-performing models evaluated at \(N=5\), the results remain disappointing.
The models correctly edit the proper code only 28\% of the time, which is insufficient for real-world usage.

\section{Discussion}
\label{sec:discussion}
\subsection{Benchmark Extension using Human-in-the-loop Feedback}

The creation of the reviewing tool (see \Cref{subsec:refinement} and \Cref{fig:human_annotation}) not only refined our evaluation pipeline but also extended the benchmark's capabilities by introducing human-centric insights and annotations. By systematically adding \texttt{human\_score} and \texttt{human\_comment} fields, the dataset now goes beyond metric-based evaluation to include nuanced qualitative assessments. These additions can enhance the interpretability and reliability of evaluation results.

Importantly, this human-augmented extension is \emph{optional} and not part of the evaluation pipeline described earlier; rather, it provides additional data that can be leveraged for further research and analysis. In particular, this enriched dataset augments the original \vtikz{} dataset with both metric outputs and structured human feedback, making it a valuable resource for fine-tuning models to better align with human judgment; conducting error analysis and failure mode discovery; exploring the space of valid code transformations. 

Moreover, the reviewing tool we developed can be reused to augment our datasets with further fine-grained human feedback. 
 Beyond static evaluation, it also enables the investigation of interactive scenarios where LLMs iteratively incorporate feedback to guide their code generation process.

\TODO{ Added for "Shallow analysis of LLM failure modes"}
\subsection{LLM failure mode}
In the following section we list reasons why LLMs fail to make the right edit to the code using the already compileable \AnnotatedVariants annotated variants. Among these annotated variants, 75 were right edits, for the remaining 225 ones, the major failure reasons were:
\begin{itemize}
    \item \textit{Feature found, wrong edit}(65) - The LLM finds what to edit but does not manage to make the right edit.
    \item \textit{Too many features edited}(19) - The LLM correctly edits the target but also alters unrelated parts, likely to maximize chances of success, but ending up making an unwanted customization.
    \item \textit{Not all feature found}(41) - The LLM only partially applies the edit; e.g., it removes a container but does not remove all the content within the container.
    \item \textit{Feature not found}(67) - The LLM does not find the feature, for example, editing the color of the wrong feature.  
    \item \textit{instruction/code not understood}(20) – Cases where the LLM fails to understand the instruction or the code. It's unclear whether the model is unsure or simply cannot apply the instruction. For example, when asked to "make the shark blue", it sometimes colors all features blue—either from misunderstanding what to color or from literal interpretation.
    \item \textit{No modifications}(13) - Happening with smaller and less performing models, that sometimes did not change anything in the original code.
\end{itemize}

\subsection{Limitations}
\paragraph{Best-Of-N measurements}
Due to the inherent randomness of LLMs, measurements with \(N=1\) at a nonzero temperature may not accurately reflect the model's average performance. To mitigate this, we also provide evaluations using \(N=5\) for the Llama-3.3-70B, DeepSeek-R1-Distill-Llama-70B, and GPT-4o models.

\paragraph{Parameterized evaluation}
The parameterization matching in \Cref{sec:param_code} uses four commands, which may not cover cases where certain lines are optional or line placement does not affect correctness. These commands were defined during an initial data exploration phase, where we analyzed variants classified as perfect by human evaluators, despite differences from the reference perfect solution. When these commands could not capture all variations, we duplicated the solution and manually created alternative variants.

\paragraph{\vtikz dataset}
The \vtikz dataset consists of 100 manually curated examples. While this is relatively small compared to datasets such as nllg/datikz, it provides high-quality, handcrafted examples representative of customization scenarios. Our evaluation shows that LLMs can successfully solve up to 28\% of these examples with \(k=5\), highlighting fundamental limitations in their capabilities and underscoring the relevance of our benchmark.

\paragraph{TikZ scope}
The benchmark targets TikZ which arguably does not require full software engineering (SE) expertise. But it provides a testbed for evaluating cross-modal consistency in code customization -- a key concern in, e.g., front-end development or end-user programming.
Although the task does not capture the full complexity of SE projects spanning multiple files and artifacts, strong performance on \vtikz would offer an encouraging signal for broader SE challenges.

\subsection{Limits and Challenges of Integrating Vision into LLM-based Code Customization}

Integrating vision modules in LLM-based code customization is a promising direction to enhance performances of LLMs in the dataset's task. This section discusses the potential benefits and challenges associated with such an approach. 

To integrate vision into LLM-based customization, several approaches can be leveraged, including Vision-Language Models (VLMs), multimodal LLMs, and object detection techniques. These methods can help assessing whether an image created from a generated code is a solution to the given instruction. In the context of an agentic system, a generated solution could be evaluated in a self-refinement protocol, however, each approach presents distinct limitations, which we examine in this section.

Using our benchmark solutions as ground truth, we conducted two preliminary tests to evaluate the potential of VLMs and multimodal models as oracles. The goal was to determine whether these models could correctly classify a reference solution as one that applies the prompt. The prompt illustrated in Figure~\ref{fig:myprompt} was used for both tests.

\begin{figure}
    \centering
    \prompt{ You are an image classification agent. Your role is to evaluate whether a given instruction has been correctly applied to an image. You are given the original image, the modified image, and an instruction. \\Response Format:

- Provide a step-by-step analysis of the image in relation to the instruction.\\
- Conclude your response with either <YES> or <NO> on a new line, depending on whether the instruction was applied.\\
- Ensure that <YES> or <NO> is enclosed within less than (<) and greater than (>) signs and appears on a separate line at the end of the response.\\
- Ensure the less than (<) and greater than (>) signs are only used at the end of the response and nowhere else.\\

Was the instruction "\{instruction\}" applied to the image? }

    \caption{Prompt used for the evaluation of VLM/Multimodal models to serve as oracles}
    \label{fig:myprompt}
\end{figure}

\paragraph{VLM limitations in TikZ code verification tasks}
In our initial test, we provided Llama3.2-Vision with correct image solutions and asked if the instruction was applied. The model incorrectly flagged 51 out of 100 instructions as not applied. However, some instructions, like "Make the cat wider," require the original image for assessment, so we conducted a second test using GPT-4o-mini with both original and edited images, in which 28 out of 100 instructions were misclassified. A similar evaluation on incorrect solutions showed that LLaMA-90B-Vision misclassified 48 out of 232, while GPT-4o-mini misclassified 20. Although these models perform better in identifying incorrect solutions, they still exhibit limitations, showing they are unreliable for TikZ code verification.

\paragraph{Challenges in using object detection for code customization}
Another approach involves leveraging object detection solutions to identify regions that should be edited. However, this approach exhibits several limitations. First, many object detection models operate either with a predefined and limited vocabulary, or through zero-shot object detection. The former cannot identify specific features, like "eyes," while the latter requires dynamic vocabulary definition, adding complexity.
 Using zero-shot detection as an oracle also presents challenges: even if regions are identified, verifying changes is difficult. Image alterations may lead to false positives, and edits outside the defined region can result in false negatives.
Solutions such as Gemini Spatial\footnote{\url{https://github.com/google-gemini/starter-applets/tree/main}} and OmniParser~\cite{lu2024omniparserpurevisionbased} offer similar capabilities but are subject to the same limitations.
 Finally, even if a perfect oracle were to exist, another challenge arises in mapping image features to the code itself. While languages like HTML allow inspecting elements to find their corresponding definitions in the code, this feature is not available in TikZ and cannot be assumed in all languages.

\section{Related Work}
\label{sec:related}

\emph{Repository-level code edit Benchmark.} 
Several benchmarks evaluate LLMs on end-to-end software development at the repository level, often leveraging existing GitHub issues, such as SWE-Bench \cite{jimenez_swe-bench_2024}, SWE-Bench+ \cite{aleithan_swe-bench_2024}, REPOCOD \cite{liang_can_2024}, and RES-Q \cite{labash_res-q_2024}. While related to our work, these benchmarks primarily assess autonomous agents executing technical, code-related instructions and do not focus on the visual output of code.


\emph{Code generation benchmarks from text instruction.} 
LLMs are widely evaluated on code generation tasks, exemplified by MBPP \cite{austin_program_2021}, HumanEval \cite{chen_evaluating_2021}, and many others \cite{zhuo_bigcodebench_2024,hendrycks_measuring_2021,wang_mconala_2023,haller_pecc_2024,jain_livecodebench_2024}. These benchmarks primarily assess full-code generation for programming challenges. Some works extend this evaluation to codes with graphical outputs, such as web frameworks \cite{cui_webapp1k_2024}, mobile/desktop applications \cite{zheng_how_2024}, plots \cite{zhuo_bigcodebench_2024,goswami_plotgen_2025} and SVG \cite{xing_svgfusion_2024,rodriguez_starvector_2024}. 
However, they focus on generating complete code rather than editing existing code. 

\emph{Code generation benchmark from multimodal instruction.} 
Certain benchmarks evaluate LLMs on code generation from multimodal inputs. HumanEval-V \cite{zhang_humaneval-v_2024} introduces images alongside textual instructions for coding challenges, Design2Code \cite{si_design2code_2024} evaluates website generation from images, and Detikzify \cite{belouadi_detikzify_2024} assesses TikZ code generation from hand-drawn sketches. Additionally, \citet{wei_words_2024} evaluates TikZ code generation from textual or visual inputs. While closely related to our work, these benchmarks assess code generation capabilities, whereas we focus on code editing.

\emph{Graphic code generation solutions.} 
Some approaches generate or edit graphical code based on high-level prompts, particularly for animations or creative coding, using custom LLM-based interfaces \cite{tseng_keyframer_2024,angert_spellburst_2023} or specialized frameworks \cite{liu_logomotion_2024}. Many works have also been conducted in the context of SVG\cite{xing_svgfusion_2024,rodriguez_starvector_2024,xing_svgdreamer_2024,wu_iconshop_2023,banerjee_svgcraft_2024}.
Other works, such as Automatikz \cite{belouadi_automatikz_2024} and Diagrammer \cite{zala_diagrammergpt_2024}, incorporate planning and self-refinement loops for structured diagram generation. 
These approaches however focused on full-code generation out of textual instructions. The customization task demands a precise understanding of both code structure and user intent to make targeted edits without breaking the original diagram. 
 For instance, \citeauthor{belouadi_automatikz_2024}~\cite{belouadi_automatikz_2024} did fine-tune a model on TikZ code and created their own model CLiMA, but our preliminary experiments revealed they did not work for code edition, generating completely different codes than the one provided.

\emph{Visual and editing code benchmarks.} 
Image generation is a well-established field, with numerous benchmarks assessing text-to-image generation \cite{huang_t2i-compbench_2023,han_evalmuse-40k_2024,bakr_hrs-bench_2023,cho_visual_2023}. However, these benchmarks are unrelated to code generation or editing, which is the focus of our contribution.
To the best of our knowledge, \citet{wei_words_2024} is the only work that partially evaluates code editing, but it does so by regenerating code from an image before applying edits, rather than directly modifying existing code. The area of code customization remains under-explored, particularly for TikZ code, which our contribution is the first to address to the best of our knowledge.

\emph{Web agents code benchmarks.} The notion of autonomous web agents is an active research area aiming to develop AIs capable of navigating and acting on the web to accomplish complex tasks in natural language~\cite{kapoor2024aiagentsmatter}. Realistic testing environments and rigorous benchmarks like WebArena~\cite{zhou2023webarena} and WebGames~\cite{thomas2025webgameschallenginggeneralpurposewebbrowsing} are essential for evaluating their progress and identifying the challenges that need to be overcome. Platforms like OpenHands  (formerly OpenDevin)~\cite{openhands} facilitate the creation and experimentation of these web agents. 
Our benchmark on TikZ provides another code-centric task that one might find on the Web for vector image creation applications. It provides a mean of evaluating the ability of future WebAgents to work on these kinds of applications. 

\vspace*{-4mm}
\section{Conclusion}
In this paper, we challenged LLMs' capabilities in customizing TikZ code from end-users instructions in natural language. This task requires accurately identifying the relevant parts of the source code to customize, and generating a syntactically correct variant that aligns with the user's intended visual outcome.
We first presented a dataset of TikZ customization scenarios, covering a wide range of edit types and complexities. This dataset serves as the backbone of the \vtikz Benchmark, which enables a systematic evaluation of state-of-the-art LLMs. As part of this benchmark, we introduced a parameterized ground-truth framework to rigorously assess the quality of generated TikZ code. Additionally, we contributed with a reviewing tool and a collection of \AnnotatedVariants annotated LLM-generated outputs. Our evaluation shows that current LLMs perform poorly in TikZ customization, achieving the desired output in only 13\% of cases in one-shot and 28\% of cases despite having 5 tries to succeed. These results highlight limitations of existing models and the need for more advanced self-refining solutions, possibly leveraging vision models and agent-based approaches. This work thus opens room for multimodal LLM-based solutions for code.

\textbf{Benchmark evolution and policy.} The results presented rely on version 1.0 of the benchmark, which is subject to future revisions. Planned updates include expanding the range of TikZ code and editing scenarios, potentially incorporating missing correct solutions, and establishing a leaderboard. A split between public and private benchmark instances is also under consideration.
\TODO{mention website here again?}

\textbf{Future work.} Multiple perspectives can be explored in the context of graphical language editing. While TikZ serves as a strong testbed, extending the evaluation to additional languages or libraries— such as SVG, P5.js, Turtle, or Matplotlib—would enhance the generalizability of our findings. 
Beyond standalone graphical languages, future research could investigate code editing capabilities in domains such as game development, mobile applications, or web interfaces, that typically involve large code bases. These applications may require evaluating more complex systems, including autonomous agents, rather than solely LLMs. 
Our findings indicate that existing solutions may be insufficient to guarantee correct visual outputs. Future work could focus on developing more advanced approaches, such as agent-based systems or iterative refinement strategies, to improve the accuracy and reliability of code edits. 

\textbf{Acknowledgements.} This work is supported by the Inria Défi LLM4Code.

\vspace{-1mm}


%


\bibliography{ref}


\begin{thebibliography}{45}


\ifx \showCODEN    \undefined \def \showCODEN     #1{\unskip}     \fi
\ifx \showDOI      \undefined \def \showDOI       #1{#1}\fi
\ifx \showISBNx    \undefined \def \showISBNx     #1{\unskip}     \fi
\ifx \showISBNxiii \undefined \def \showISBNxiii  #1{\unskip}     \fi
\ifx \showISSN     \undefined \def \showISSN      #1{\unskip}     \fi
\ifx \showLCCN     \undefined \def \showLCCN      #1{\unskip}     \fi
\ifx \shownote     \undefined \def \shownote      #1{#1}          \fi
\ifx \showarticletitle \undefined \def \showarticletitle #1{#1}   \fi
\ifx \showURL      \undefined \def \showURL       {\relax}        \fi
\providecommand\bibfield[2]{#2}
\providecommand\bibinfo[2]{#2}
\providecommand\natexlab[1]{#1}
\providecommand\showeprint[2][]{arXiv:#2}

\bibitem[Acher et~al\mbox{.}(2023)]%
        {acher_programming_2023}
\bibfield{author}{\bibinfo{person}{Mathieu Acher},
  \bibinfo{person}{Jos{\'e}~Galindo Duarte}, {and} \bibinfo{person}{Jean-Marc
  J{\'e}z{\'e}quel}.} \bibinfo{year}{2023}\natexlab{}.
\newblock \showarticletitle{On {Programming} {Variability} with {Large}
  {Language} {Model}-based {Assistant}} \emph{(\bibinfo{series}{{SPLC} '23},
  Vol.~\bibinfo{volume}{A})}. \bibinfo{pages}{8--14}.
\newblock
\showISBNx{979-8-4007-0091-0}
\urldef\tempurl%
\url{https://dl.acm.org/doi/10.1145/3579027.3608972}
\showURL{%
\tempurl}


\bibitem[Aleithan et~al\mbox{.}(2024)]%
        {aleithan_swe-bench_2024}
\bibfield{author}{\bibinfo{person}{Reem Aleithan}, \bibinfo{person}{Haoran
  Xue}, \bibinfo{person}{Mohammad~Mahdi Mohajer}, \bibinfo{person}{Elijah
  Nnorom}, \bibinfo{person}{Gias Uddin}, {and} \bibinfo{person}{Song Wang}.}
  \bibinfo{year}{2024}\natexlab{}.
\newblock \bibinfo{title}{{SWE}-{Bench}+: {Enhanced} {Coding} {Benchmark} for
  {LLMs}}.
\newblock
\newblock
\newblock
\shownote{arXiv:2410.06992 version: 1}.


\bibitem[Angert et~al\mbox{.}(2023)]%
        {angert_spellburst_2023}
\bibfield{author}{\bibinfo{person}{Tyler Angert}, \bibinfo{person}{Miroslav
  Suzara}, \bibinfo{person}{Jenny Han}, \bibinfo{person}{Christopher Pondoc},
  {and} \bibinfo{person}{Hariharan Subramonyam}.}
  \bibinfo{year}{2023}\natexlab{}.
\newblock \showarticletitle{Spellburst: {A} {Node}-based {Interface} for
  {Exploratory} {Creative} {Coding} with {Natural} {Language} {Prompts}}
  \emph{(\bibinfo{series}{{UIST} '23})}. \bibinfo{address}{New York, NY, USA},
  \bibinfo{pages}{1--22}.
\newblock
\showISBNx{979-8-4007-0132-0}


\bibitem[Austin et~al\mbox{.}(2021)]%
        {austin_program_2021}
\bibfield{author}{\bibinfo{person}{Jacob Austin}, \bibinfo{person}{Augustus
  Odena}, \bibinfo{person}{Maxwell Nye}, \bibinfo{person}{Maarten Bosma},
  \bibinfo{person}{Henryk Michalewski}, \bibinfo{person}{David Dohan},
  \bibinfo{person}{Ellen Jiang}, \bibinfo{person}{Carrie Cai},
  \bibinfo{person}{Michael Terry}, \bibinfo{person}{Quoc Le}, {and}
  \bibinfo{person}{Charles Sutton}.} \bibinfo{year}{2021}\natexlab{}.
\newblock \bibinfo{title}{Program {Synthesis} with {Large} {Language}
  {Models}}.
\newblock
\newblock
\newblock
\shownote{arXiv:2108.07732}.


\bibitem[Bakr et~al\mbox{.}(2023)]%
        {bakr_hrs-bench_2023}
\bibfield{author}{\bibinfo{person}{Eslam~Mohamed Bakr},
  \bibinfo{person}{Pengzhan Sun}, \bibinfo{person}{Xiaoqian Shen},
  \bibinfo{person}{Faizan~Farooq Khan}, \bibinfo{person}{Li~Erran Li}, {and}
  \bibinfo{person}{Mohamed Elhoseiny}.} \bibinfo{year}{2023}\natexlab{}.
\newblock \bibinfo{title}{{HRS}-{Bench}: {Holistic}, {Reliable} and {Scalable}
  {Benchmark} for {Text}-to-{Image} {Models}}.
\newblock
\newblock
\newblock
\shownote{arXiv:2304.05390}.


\bibitem[Banerjee et~al\mbox{.}(2024)]%
        {banerjee_svgcraft_2024}
\bibfield{author}{\bibinfo{person}{Ayan Banerjee}, \bibinfo{person}{Nityanand
  Mathur}, \bibinfo{person}{Josep Llad{\'o}s}, \bibinfo{person}{Umapada Pal},
  {and} \bibinfo{person}{Anjan Dutta}.} \bibinfo{year}{2024}\natexlab{}.
\newblock \bibinfo{title}{{SVGCraft}: {Beyond} {Single} {Object}
  {Text}-to-{SVG} {Synthesis} with {Comprehensive} {Canvas} {Layout}}.
\newblock
\newblock
\newblock
\shownote{arXiv:2404.00412}.


\bibitem[Belouadi et~al\mbox{.}(2024a)]%
        {belouadi_automatikz_2024}
\bibfield{author}{\bibinfo{person}{Jonas Belouadi}, \bibinfo{person}{Anne
  Lauscher}, {and} \bibinfo{person}{Steffen Eger}.}
  \bibinfo{year}{2024}\natexlab{a}.
\newblock \bibinfo{title}{{AutomaTikZ}: {Text}-{Guided} {Synthesis} of
  {Scientific} {Vector} {Graphics} with {TikZ}}.
\newblock
\newblock
\newblock
\shownote{arXiv:2310.00367 [cs]}.


\bibitem[Belouadi et~al\mbox{.}(2024b)]%
        {belouadi_detikzify_2024}
\bibfield{author}{\bibinfo{person}{Jonas Belouadi},
  \bibinfo{person}{Simone~Paolo Ponzetto}, {and} \bibinfo{person}{Steffen
  Eger}.} \bibinfo{year}{2024}\natexlab{b}.
\newblock \bibinfo{title}{{DeTikZify}: {Synthesizing} {Graphics} {Programs} for
  {Scientific} {Figures} and {Sketches} with {TikZ}}.
\newblock
\newblock
\newblock
\shownote{arXiv:2405.15306 [cs]}.


\bibitem[Bubeck et~al\mbox{.}(2023)]%
        {bubeck_sparks_2023}
\bibfield{author}{\bibinfo{person}{S{\'e}bastien Bubeck},
  \bibinfo{person}{Varun Chandrasekaran}, \bibinfo{person}{Ronen Eldan},
  \bibinfo{person}{Johannes Gehrke}, \bibinfo{person}{Eric Horvitz},
  \bibinfo{person}{Ece Kamar}, \bibinfo{person}{Peter Lee},
  \bibinfo{person}{Yin~Tat Lee}, \bibinfo{person}{Yuanzhi Li},
  \bibinfo{person}{Scott Lundberg}, \bibinfo{person}{Harsha Nori},
  \bibinfo{person}{Hamid Palangi}, \bibinfo{person}{Marco~Tulio Ribeiro}, {and}
  \bibinfo{person}{Yi Zhang}.} \bibinfo{year}{2023}\natexlab{}.
\newblock \bibinfo{title}{Sparks of {Artificial} {General} {Intelligence}:
  {Early} experiments with {GPT}-4}.
\newblock
\newblock
\newblock
\shownote{arXiv:2303.12712}.


\bibitem[Chen and al.(2021)]%
        {chen_evaluating_2021}
\bibfield{author}{\bibinfo{person}{Mark Chen} {and} \bibinfo{person}{al.}}
  \bibinfo{year}{2021}\natexlab{}.
\newblock \bibinfo{title}{Evaluating {Large} {Language} {Models} {Trained} on
  {Code}}.
\newblock
\newblock
\newblock
\shownote{arXiv:2107.03374}.


\bibitem[Cho et~al\mbox{.}(2023)]%
        {cho_visual_2023}
\bibfield{author}{\bibinfo{person}{Jaemin Cho}, \bibinfo{person}{Abhay Zala},
  {and} \bibinfo{person}{Mohit Bansal}.} \bibinfo{year}{2023}\natexlab{}.
\newblock \bibinfo{title}{Visual {Programming} for {Text}-to-{Image}
  {Generation} and {Evaluation}}.
\newblock
\newblock
\newblock
\shownote{arXiv:2305.15328}.


\bibitem[Cui(2024)]%
        {cui_webapp1k_2024}
\bibfield{author}{\bibinfo{person}{Yi Cui}.} \bibinfo{year}{2024}\natexlab{}.
\newblock \bibinfo{title}{{WebApp1K}: {A} {Practical} {Code}-{Generation}
  {Benchmark} for {Web} {App} {Development}}.
\newblock
\newblock
\newblock
\shownote{arXiv:2408.00019 version: 1}.


\bibitem[Eghbali and Pradel(2023)]%
        {eghbali_crystalbleu_2023}
\bibfield{author}{\bibinfo{person}{Aryaz Eghbali} {and}
  \bibinfo{person}{Michael Pradel}.} \bibinfo{year}{2023}\natexlab{}.
\newblock \showarticletitle{{CrystalBLEU}: {Precisely} and {Efficiently}
  {Measuring} the {Similarity} of {Code}} \emph{(\bibinfo{series}{{ASE} '22})}.
  \bibinfo{address}{New York, NY, USA}, \bibinfo{pages}{1--12}.
\newblock
\showISBNx{978-1-4503-9475-8}


\bibitem[Goswami et~al\mbox{.}(2025)]%
        {goswami_plotgen_2025}
\bibfield{author}{\bibinfo{person}{Kanika Goswami}, \bibinfo{person}{Puneet
  Mathur}, \bibinfo{person}{Ryan Rossi}, {and} \bibinfo{person}{Franck
  Dernoncourt}.} \bibinfo{year}{2025}\natexlab{}.
\newblock \bibinfo{title}{{PlotGen}: {Multi}-{Agent} {LLM}-based {Scientific}
  {Data} {Visualization} via {Multimodal} {Feedback}}.
\newblock
\newblock
\newblock
\shownote{arXiv:2502.00988 version: 1}.


\bibitem[Haller et~al\mbox{.}(2024)]%
        {haller_pecc_2024}
\bibfield{author}{\bibinfo{person}{Patrick Haller}, \bibinfo{person}{Jonas
  Golde}, {and} \bibinfo{person}{Alan Akbik}.} \bibinfo{year}{2024}\natexlab{}.
\newblock \bibinfo{title}{{PECC}: {Problem} {Extraction} and {Coding}
  {Challenges}}.
\newblock
\newblock
\newblock
\shownote{arXiv:2404.18766}.


\bibitem[Han et~al\mbox{.}(2024)]%
        {han_evalmuse-40k_2024}
\bibfield{author}{\bibinfo{person}{Shuhao Han}, \bibinfo{person}{Haotian Fan},
  \bibinfo{person}{Jiachen Fu}, \bibinfo{person}{Liang Li},
  \bibinfo{person}{Tao Li}, \bibinfo{person}{Junhui Cui},
  \bibinfo{person}{Yunqiu Wang}, \bibinfo{person}{Yang Tai},
  \bibinfo{person}{Jingwei Sun}, \bibinfo{person}{Chunle Guo}, {and}
  \bibinfo{person}{Chongyi Li}.} \bibinfo{year}{2024}\natexlab{}.
\newblock \bibinfo{title}{{EvalMuse}-{40K}: {A} {Reliable} and {Fine}-{Grained}
  {Benchmark} with {Comprehensive} {Human} {Annotations} for {Text}-to-{Image}
  {Generation} {Model} {Evaluation}}.
\newblock
\newblock
\newblock
\shownote{arXiv:2412.18150}.


\bibitem[Hendrycks et~al\mbox{.}(2021)]%
        {hendrycks_measuring_2021}
\bibfield{author}{\bibinfo{person}{Dan Hendrycks}, \bibinfo{person}{Steven
  Basart}, \bibinfo{person}{Saurav Kadavath}, \bibinfo{person}{Mantas Mazeika},
  \bibinfo{person}{Akul Arora}, \bibinfo{person}{Ethan Guo},
  \bibinfo{person}{Collin Burns}, \bibinfo{person}{Samir Puranik},
  \bibinfo{person}{Horace He}, \bibinfo{person}{Dawn Song}, {and}
  \bibinfo{person}{Jacob Steinhardt}.} \bibinfo{year}{2021}\natexlab{}.
\newblock \bibinfo{title}{Measuring {Coding} {Challenge} {Competence} {With}
  {APPS}}.
\newblock
\newblock
\newblock
\shownote{arXiv:2105.09938}.


\bibitem[Hou et~al\mbox{.}(2024)]%
        {hou_large_2024}
\bibfield{author}{\bibinfo{person}{Xinyi Hou}, \bibinfo{person}{Yanjie Zhao},
  \bibinfo{person}{Yue Liu}, \bibinfo{person}{Zhou Yang},
  \bibinfo{person}{Kailong Wang}, \bibinfo{person}{Li Li},
  \bibinfo{person}{Xiapu Luo}, \bibinfo{person}{David Lo},
  \bibinfo{person}{John Grundy}, {and} \bibinfo{person}{Haoyu Wang}.}
  \bibinfo{year}{2024}\natexlab{}.
\newblock \showarticletitle{Large {Language} {Models} for {Software}
  {Engineering}: {A} {Systematic} {Literature} {Review}}.
\newblock \bibinfo{journal}{\emph{ACM Trans. Softw. Eng. Methodol.}}
  \bibinfo{volume}{33}, \bibinfo{number}{8} (\bibinfo{date}{Dec.}
  \bibinfo{year}{2024}), \bibinfo{pages}{220:1--220:79}.
\newblock
\showISSN{1049-331X}
\urldef\tempurl%
\url{https://dl.acm.org/doi/10.1145/3695988}
\showURL{%
\tempurl}


\bibitem[Huang et~al\mbox{.}(2023)]%
        {huang_t2i-compbench_2023}
\bibfield{author}{\bibinfo{person}{Kaiyi Huang}, \bibinfo{person}{Kaiyue Sun},
  \bibinfo{person}{Enze Xie}, \bibinfo{person}{Zhenguo Li}, {and}
  \bibinfo{person}{Xihui Liu}.} \bibinfo{year}{2023}\natexlab{}.
\newblock \bibinfo{title}{{T2I}-{CompBench}: {A} {Comprehensive} {Benchmark}
  for {Open}-world {Compositional} {Text}-to-image {Generation}}.
\newblock
\newblock
\newblock
\shownote{arXiv:2307.06350}.


\bibitem[Jain et~al\mbox{.}(2024)]%
        {jain_livecodebench_2024}
\bibfield{author}{\bibinfo{person}{Naman Jain}, \bibinfo{person}{King Han},
  \bibinfo{person}{Alex Gu}, \bibinfo{person}{Wen-Ding Li},
  \bibinfo{person}{Fanjia Yan}, \bibinfo{person}{Tianjun Zhang},
  \bibinfo{person}{Sida Wang}, \bibinfo{person}{Armando Solar-Lezama},
  \bibinfo{person}{Koushik Sen}, {and} \bibinfo{person}{Ion Stoica}.}
  \bibinfo{year}{2024}\natexlab{}.
\newblock \bibinfo{title}{{LiveCodeBench}: {Holistic} and {Contamination}
  {Free} {Evaluation} of {Large} {Language} {Models} for {Code}}.
\newblock
\newblock
\newblock
\shownote{arXiv:2403.07974}.


\bibitem[Jimenez et~al\mbox{.}(2024)]%
        {jimenez_swe-bench_2024}
\bibfield{author}{\bibinfo{person}{Carlos~E. Jimenez}, \bibinfo{person}{John
  Yang}, \bibinfo{person}{Alexander Wettig}, \bibinfo{person}{Shunyu Yao},
  \bibinfo{person}{Kexin Pei}, \bibinfo{person}{Ofir Press}, {and}
  \bibinfo{person}{Karthik Narasimhan}.} \bibinfo{year}{2024}\natexlab{}.
\newblock \bibinfo{title}{{SWE}-bench: {Can} {Language} {Models} {Resolve}
  {Real}-{World} {GitHub} {Issues}?}
\newblock
\newblock
\newblock
\shownote{arXiv:2310.06770}.


\bibitem[Kapoor et~al\mbox{.}(2024)]%
        {kapoor2024aiagentsmatter}
\bibfield{author}{\bibinfo{person}{Sayash Kapoor}, \bibinfo{person}{Benedikt
  Stroebl}, \bibinfo{person}{Zachary~S. Siegel}, \bibinfo{person}{Nitya
  Nadgir}, {and} \bibinfo{person}{Arvind Narayanan}.}
  \bibinfo{year}{2024}\natexlab{}.
\newblock \bibinfo{title}{{AI} {Agents} {That} {Matter}}.
\newblock
\newblock
\newblock
\shownote{arXiv:2407.01502 [cs]}.


\bibitem[LaBash et~al\mbox{.}(2024)]%
        {labash_res-q_2024}
\bibfield{author}{\bibinfo{person}{Beck LaBash}, \bibinfo{person}{August
  Rosedale}, \bibinfo{person}{Alex Reents}, \bibinfo{person}{Lucas Negritto},
  {and} \bibinfo{person}{Colin Wiel}.} \bibinfo{year}{2024}\natexlab{}.
\newblock \bibinfo{title}{{RES}-{Q}: {Evaluating} {Code}-{Editing} {Large}
  {Language} {Model} {Systems} at the {Repository} {Scale}}.
\newblock
\newblock
\newblock
\shownote{arXiv:2406.16801}.


\bibitem[Liang et~al\mbox{.}(2024)]%
        {liang_can_2024}
\bibfield{author}{\bibinfo{person}{Shanchao Liang}, \bibinfo{person}{Yiran Hu},
  \bibinfo{person}{Nan Jiang}, {and} \bibinfo{person}{Lin Tan}.}
  \bibinfo{year}{2024}\natexlab{}.
\newblock \bibinfo{title}{Can {Language} {Models} {Replace} {Programmers}?
  {REPOCOD} {Says} '{Not} {Yet}'}.
\newblock
\newblock
\newblock
\shownote{arXiv:2410.21647}.


\bibitem[Liu et~al\mbox{.}(2024c)]%
        {liu_large_2024}
\bibfield{author}{\bibinfo{person}{Junwei Liu}, \bibinfo{person}{Kaixin Wang},
  \bibinfo{person}{Yixuan Chen}, \bibinfo{person}{Xin Peng},
  \bibinfo{person}{Zhenpeng Chen}, \bibinfo{person}{Lingming Zhang}, {and}
  \bibinfo{person}{Yiling Lou}.} \bibinfo{year}{2024}\natexlab{c}.
\newblock \bibinfo{title}{Large {Language} {Model}-{Based} {Agents} for
  {Software} {Engineering}: {A} {Survey}}.
\newblock
\newblock
\newblock
\shownote{arXiv:2409.02977}.


\bibitem[Liu et~al\mbox{.}(2024b)]%
        {liu_logomotion_2024}
\bibfield{author}{\bibinfo{person}{Vivian Liu}, \bibinfo{person}{Rubaiat~Habib
  Kazi}, \bibinfo{person}{Li-Yi Wei}, \bibinfo{person}{Matthew Fisher},
  \bibinfo{person}{Timothy Langlois}, \bibinfo{person}{Seth Walker}, {and}
  \bibinfo{person}{Lydia Chilton}.} \bibinfo{year}{2024}\natexlab{b}.
\newblock \bibinfo{title}{{LogoMotion}: {Visually} {Grounded} {Code}
  {Generation} for {Content}-{Aware} {Animation}}.
\newblock
\newblock
\newblock
\shownote{arXiv:2405.07065}.


\bibitem[Liu et~al\mbox{.}(2024a)]%
        {liu_empirical_2024}
\bibfield{author}{\bibinfo{person}{Yongkun Liu}, \bibinfo{person}{Jiachi Chen},
  \bibinfo{person}{Tingting Bi}, \bibinfo{person}{John Grundy},
  \bibinfo{person}{Yanlin Wang}, \bibinfo{person}{Jianxing Yu},
  \bibinfo{person}{Ting Chen}, \bibinfo{person}{Yutian Tang}, {and}
  \bibinfo{person}{Zibin Zheng}.} \bibinfo{year}{2024}\natexlab{a}.
\newblock \bibinfo{title}{An {Empirical} {Study} on {Low} {Code} {Programming}
  using {Traditional} vs {Large} {Language} {Model} {Support}}.
\newblock
\newblock
\newblock
\shownote{arXiv:2402.01156}.


\bibitem[Lu et~al\mbox{.}(2024)]%
        {lu2024omniparserpurevisionbased}
\bibfield{author}{\bibinfo{person}{Yadong Lu}, \bibinfo{person}{Jianwei Yang},
  \bibinfo{person}{Yelong Shen}, {and} \bibinfo{person}{Ahmed Awadallah}.}
  \bibinfo{year}{2024}\natexlab{}.
\newblock \bibinfo{title}{OmniParser for Pure Vision Based GUI Agent}.
\newblock
\newblock
\showeprint[arxiv]{2408.00203}~[cs.CV]
\newblock
\shownote{arXiv:2408.00203}.


\bibitem[Rodriguez et~al\mbox{.}(2024)]%
        {rodriguez_starvector_2024}
\bibfield{author}{\bibinfo{person}{Juan~A. Rodriguez}, \bibinfo{person}{Abhay
  Puri}, \bibinfo{person}{Shubham Agarwal}, \bibinfo{person}{Issam~H. Laradji},
  \bibinfo{person}{Pau Rodriguez}, \bibinfo{person}{Sai Rajeswar},
  \bibinfo{person}{David Vazquez}, \bibinfo{person}{Christopher Pal}, {and}
  \bibinfo{person}{Marco Pedersoli}.} \bibinfo{year}{2024}\natexlab{}.
\newblock \bibinfo{title}{{StarVector}: {Generating} {Scalable} {Vector}
  {Graphics} {Code} from {Images} and {Text}}.
\newblock
\newblock
\newblock
\shownote{arXiv:2312.11556}.


\bibitem[Si et~al\mbox{.}(2024)]%
        {si_design2code_2024}
\bibfield{author}{\bibinfo{person}{Chenglei Si}, \bibinfo{person}{Yanzhe
  Zhang}, \bibinfo{person}{Zhengyuan Yang}, \bibinfo{person}{Ruibo Liu}, {and}
  \bibinfo{person}{Diyi Yang}.} \bibinfo{year}{2024}\natexlab{}.
\newblock \bibinfo{title}{{Design2Code}: {How} {Far} {Are} {We} {From}
  {Automating} {Front}-{End} {Engineering}?}
\newblock
\newblock
\newblock
\shownote{arXiv:2403.03163}.


\bibitem[Thomas et~al\mbox{.}(2025)]%
        {thomas2025webgameschallenginggeneralpurposewebbrowsing}
\bibfield{author}{\bibinfo{person}{George Thomas}, \bibinfo{person}{Alex~J.
  Chan}, \bibinfo{person}{Jikun Kang}, \bibinfo{person}{Wenqi Wu},
  \bibinfo{person}{Filippos Christianos}, \bibinfo{person}{Fraser Greenlee},
  \bibinfo{person}{Andy Toulis}, {and} \bibinfo{person}{Marvin Purtorab}.}
  \bibinfo{year}{2025}\natexlab{}.
\newblock \bibinfo{title}{{WebGames}: {Challenging} {General}-{Purpose}
  {Web}-{Browsing} {AI} {Agents}}.
\newblock
\newblock
\newblock
\shownote{arXiv:2502.18356 [cs]}.


\bibitem[Tseng et~al\mbox{.}(2024)]%
        {tseng_keyframer_2024}
\bibfield{author}{\bibinfo{person}{Tiffany Tseng}, \bibinfo{person}{Ruijia
  Cheng}, {and} \bibinfo{person}{Jeffrey Nichols}.}
  \bibinfo{year}{2024}\natexlab{}.
\newblock \bibinfo{title}{Keyframer: {Empowering} {Animation} {Design} using
  {Large} {Language} {Models}}.
\newblock
\newblock
\newblock
\shownote{arXiv:2402.06071}.


\bibitem[Wang et~al\mbox{.}(2025)]%
        {openhands}
\bibfield{author}{\bibinfo{person}{Xingyao Wang}, \bibinfo{person}{Boxuan Li},
  \bibinfo{person}{Yufan Song}, \bibinfo{person}{Frank~F. Xu},
  \bibinfo{person}{Xiangru Tang}, \bibinfo{person}{Mingchen Zhuge},
  \bibinfo{person}{Jiayi Pan}, \bibinfo{person}{Yueqi Song},
  \bibinfo{person}{Bowen Li}, \bibinfo{person}{Jaskirat Singh},
  \bibinfo{person}{Hoang~H. Tran}, \bibinfo{person}{Fuqiang Li},
  \bibinfo{person}{Ren Ma}, \bibinfo{person}{Mingzhang Zheng},
  \bibinfo{person}{Bill Qian}, \bibinfo{person}{Yanjun Shao},
  \bibinfo{person}{Niklas Muennighoff}, \bibinfo{person}{Yizhe Zhang},
  \bibinfo{person}{Binyuan Hui}, \bibinfo{person}{Junyang Lin},
  \bibinfo{person}{Robert Brennan}, \bibinfo{person}{Hao Peng},
  \bibinfo{person}{Heng Ji}, {and} \bibinfo{person}{Graham Neubig}.}
  \bibinfo{year}{2025}\natexlab{}.
\newblock \bibinfo{title}{{OpenHands}: {An} {Open} {Platform} for {AI}
  {Software} {Developers} as {Generalist} {Agents}}.
\newblock
\newblock
\newblock
\shownote{arXiv:2407.16741 [cs]}.


\bibitem[Wang et~al\mbox{.}(2023)]%
        {wang_mconala_2023}
\bibfield{author}{\bibinfo{person}{Zhiruo Wang}, \bibinfo{person}{Grace
  Cuenca}, \bibinfo{person}{Shuyan Zhou}, \bibinfo{person}{Frank~F. Xu}, {and}
  \bibinfo{person}{Graham Neubig}.} \bibinfo{year}{2023}\natexlab{}.
\newblock \bibinfo{title}{{MCoNaLa}: {A} {Benchmark} for {Code} {Generation}
  from {Multiple} {Natural} {Languages}}.
\newblock
\newblock
\newblock
\shownote{arXiv:2203.08388}.


\bibitem[Wei et~al\mbox{.}(2024)]%
        {wei_words_2024}
\bibfield{author}{\bibinfo{person}{Jingxuan Wei}, \bibinfo{person}{Cheng Tan},
  \bibinfo{person}{Qi Chen}, \bibinfo{person}{Gaowei Wu},
  \bibinfo{person}{Siyuan Li}, \bibinfo{person}{Zhangyang Gao},
  \bibinfo{person}{Linzhuang Sun}, \bibinfo{person}{Bihui Yu}, {and}
  \bibinfo{person}{Ruifeng Guo}.} \bibinfo{year}{2024}\natexlab{}.
\newblock \bibinfo{title}{From {Words} to {Structured} {Visuals}: {A}
  {Benchmark} and {Framework} for {Text}-to-{Diagram} {Generation} and
  {Editing}}.
\newblock
\newblock
\newblock
\shownote{arXiv:2411.11916}.


\bibitem[Willison(2025)]%
        {willison_notes_2025}
\bibfield{author}{\bibinfo{person}{Simon Willison}.}
  \bibinfo{year}{2025}\natexlab{}.
\newblock \bibinfo{title}{Notes on {Google}{\textquoteright}s {Gemma} 3}.
\newblock
\newblock
\urldef\tempurl%
\url{https://simonwillison.net/2025/Mar/12/gemma-3/}
\showURL{%
\tempurl}


\bibitem[Wu et~al\mbox{.}(2023)]%
        {wu_iconshop_2023}
\bibfield{author}{\bibinfo{person}{Ronghuan Wu}, \bibinfo{person}{Wanchao Su},
  \bibinfo{person}{Kede Ma}, {and} \bibinfo{person}{Jing Liao}.}
  \bibinfo{year}{2023}\natexlab{}.
\newblock \showarticletitle{{IconShop}: {Text}-{Guided} {Vector} {Icon}
  {Synthesis} with {Autoregressive} {Transformers}}.
\newblock \bibinfo{journal}{\emph{ACM Trans. Graph.}} \bibinfo{volume}{42},
  \bibinfo{number}{6} (\bibinfo{date}{Dec.} \bibinfo{year}{2023}),
  \bibinfo{pages}{230:1--230:14}.
\newblock
\showISSN{0730-0301}


\bibitem[Xing et~al\mbox{.}(2024a)]%
        {xing_svgfusion_2024}
\bibfield{author}{\bibinfo{person}{Ximing Xing}, \bibinfo{person}{Juncheng Hu},
  \bibinfo{person}{Jing Zhang}, \bibinfo{person}{Dong Xu}, {and}
  \bibinfo{person}{Qian Yu}.} \bibinfo{year}{2024}\natexlab{a}.
\newblock \bibinfo{title}{{SVGFusion}: {Scalable} {Text}-to-{SVG} {Generation}
  via {Vector} {Space} {Diffusion}}.
\newblock
\newblock
\newblock
\shownote{arXiv:2412.10437}.


\bibitem[Xing et~al\mbox{.}(2024b)]%
        {xing_svgdreamer_2024}
\bibfield{author}{\bibinfo{person}{Ximing Xing}, \bibinfo{person}{Haitao Zhou},
  \bibinfo{person}{Chuang Wang}, \bibinfo{person}{Jing Zhang},
  \bibinfo{person}{Dong Xu}, {and} \bibinfo{person}{Qian Yu}.}
  \bibinfo{year}{2024}\natexlab{b}.
\newblock \bibinfo{title}{{SVGDreamer}: {Text} {Guided} {SVG} {Generation} with
  {Diffusion} {Model}}.
\newblock
\newblock
\newblock
\shownote{arXiv:2312.16476}.


\bibitem[Zala et~al\mbox{.}(2024)]%
        {zala_diagrammergpt_2024}
\bibfield{author}{\bibinfo{person}{Abhay Zala}, \bibinfo{person}{Han Lin},
  \bibinfo{person}{Jaemin Cho}, {and} \bibinfo{person}{Mohit Bansal}.}
  \bibinfo{year}{2024}\natexlab{}.
\newblock \bibinfo{title}{{DiagrammerGPT}: {Generating} {Open}-{Domain},
  {Open}-{Platform} {Diagrams} via {LLM} {Planning}}.
\newblock
\newblock
\newblock
\shownote{arXiv:2310.12128}.


\bibitem[Zhang et~al\mbox{.}(2024)]%
        {zhang_humaneval-v_2024}
\bibfield{author}{\bibinfo{person}{Fengji Zhang}, \bibinfo{person}{Linquan Wu},
  \bibinfo{person}{Huiyu Bai}, \bibinfo{person}{Guancheng Lin},
  \bibinfo{person}{Xiao Li}, \bibinfo{person}{Xiao Yu}, \bibinfo{person}{Yue
  Wang}, \bibinfo{person}{Bei Chen}, {and} \bibinfo{person}{Jacky Keung}.}
  \bibinfo{year}{2024}\natexlab{}.
\newblock \bibinfo{title}{{HumanEval}-{V}: {Evaluating} {Visual}
  {Understanding} and {Reasoning} {Abilities} of {Large} {Multimodal} {Models}
  {Through} {Coding} {Tasks}}.
\newblock
\newblock
\newblock
\shownote{arXiv:2410.12381}.


\bibitem[Zhang and Shasha(1989)]%
        {zhang_simple_1989}
\bibfield{author}{\bibinfo{person}{Kaizhong Zhang} {and}
  \bibinfo{person}{Dennis Shasha}.} \bibinfo{year}{1989}\natexlab{}.
\newblock \showarticletitle{Simple {Fast} {Algorithms} for the {Editing}
  {Distance} between {Trees} and {Related} {Problems}}.
\newblock \bibinfo{journal}{\emph{SIAM J. Comput.}} \bibinfo{volume}{18},
  \bibinfo{number}{6} (\bibinfo{date}{Dec.} \bibinfo{year}{1989}),
  \bibinfo{pages}{1245--1262}.
\newblock
\showISSN{0097-5397}
\newblock
\shownote{Publisher: Society for Industrial and Applied Mathematics}.


\bibitem[Zheng et~al\mbox{.}(2024)]%
        {zheng_how_2024}
\bibfield{author}{\bibinfo{person}{Dewu Zheng}, \bibinfo{person}{Yanlin Wang},
  \bibinfo{person}{Ensheng Shi}, \bibinfo{person}{Hongyu Zhang}, {and}
  \bibinfo{person}{Zibin Zheng}.} \bibinfo{year}{2024}\natexlab{}.
\newblock \bibinfo{title}{How {Well} {Do} {LLMs} {Generate} {Code} for
  {Different} {Application} {Domains}? {Benchmark} and {Evaluation}}.
\newblock
\newblock
\newblock
\shownote{arXiv:2412.18573 version: 1}.


\bibitem[Zhou et~al\mbox{.}(2023)]%
        {zhou2023webarena}
\bibfield{author}{\bibinfo{person}{Shuyan Zhou}, \bibinfo{person}{Frank~F Xu},
  \bibinfo{person}{Hao Zhu}, \bibinfo{person}{Xuhui Zhou},
  \bibinfo{person}{Robert Lo}, \bibinfo{person}{Abishek Sridhar},
  \bibinfo{person}{Xianyi Cheng}, \bibinfo{person}{Yonatan Bisk},
  \bibinfo{person}{Daniel Fried}, \bibinfo{person}{Uri Alon}, {et~al\mbox{.}}}
  \bibinfo{year}{2023}\natexlab{}.
\newblock \showarticletitle{WebArena: A Realistic Web Environment for Building
  Autonomous Agents}.
\newblock \bibinfo{journal}{\emph{arXiv preprint arXiv:2307.13854}}
  (\bibinfo{year}{2023}).
\newblock
\urldef\tempurl%
\url{https://webarena.dev}
\showURL{%
\tempurl}


\bibitem[Zhuo et~al\mbox{.}(2024)]%
        {zhuo_bigcodebench_2024}
\bibfield{author}{\bibinfo{person}{Terry~Yue Zhuo}, \bibinfo{person}{Minh~Chien
  Vu}, \bibinfo{person}{Jenny Chim}, \bibinfo{person}{Han Hu},
  \bibinfo{person}{Wenhao Yu}, \bibinfo{person}{Ratnadira Widyasari},
  \bibinfo{person}{Imam Nur~Bani Yusuf}, \bibinfo{person}{Haolan Zhan},
  \bibinfo{person}{Junda He}, \bibinfo{person}{Indraneil Paul},
  \bibinfo{person}{Simon Brunner}, \bibinfo{person}{Chen Gong},
  \bibinfo{person}{Thong Hoang}, \bibinfo{person}{Armel~Randy Zebaze},
  \bibinfo{person}{Xiaoheng Hong}, \bibinfo{person}{Wen-Ding Li},
  \bibinfo{person}{Jean Kaddour}, \bibinfo{person}{Ming Xu},
  \bibinfo{person}{Zhihan Zhang}, \bibinfo{person}{Prateek Yadav},
  \bibinfo{person}{Naman Jain}, \bibinfo{person}{Alex Gu},
  \bibinfo{person}{Zhoujun Cheng}, \bibinfo{person}{Jiawei Liu},
  \bibinfo{person}{Qian Liu}, \bibinfo{person}{Zijian Wang},
  \bibinfo{person}{David Lo}, \bibinfo{person}{Binyuan Hui},
  \bibinfo{person}{Niklas Muennighoff}, \bibinfo{person}{Daniel Fried},
  \bibinfo{person}{Xiaoning Du}, \bibinfo{person}{Harm~de Vries}, {and}
  \bibinfo{person}{Leandro~Von Werra}.} \bibinfo{year}{2024}\natexlab{}.
\newblock \bibinfo{title}{{BigCodeBench}: {Benchmarking} {Code} {Generation}
  with {Diverse} {Function} {Calls} and {Complex} {Instructions}}.
\newblock
\newblock
\newblock
\shownote{arXiv:2406.15877}.


\end{thebibliography}


\end{document}